# GSEP: A ROBUST VOCAL AND ACCOMPANIMENT SEPARATION SYSTEM USING GATED CBHG MODULE AND LOUDNESS NORMALIZATION

*Soochul Park and Ben Sangbae Chon*

Gaudio Lab, Inc.

## ABSTRACT

In the field of audio signal processing research, source separation has been a popular research topic for a long time and the recent adoption of the deep neural networks have shown a significant improvement in performance. The improvement vitalizes the industry to productize audio deep learning based products and services including Karaoke in the music streaming apps and dialogue enhancement in the UHDTV. For these early markets, we defined a set of design principles of the vocal and accompaniment separation model in terms of robustness, quality, and cost. In this paper, we introduce GSEP (Gaudio source SEParation system), a robust vocal and accompaniment separation system using a Gated-CBHG module, mask warping, and loudness normalization and it was verified that the proposed system satisfies all three principles and outperforms the state-of-the-art systems both in objective measure and subjective assessment through experiments.

***Index Terms***— audio source separation, gated-CBHG, loudness normalization, dialogue enhancement, Karaoke

## 1. INTRODUCTION

Recent source separation technologies based on deep learning has shown a significant improvement in performance and new products are being introduced in the market - Music streaming services like Spotify, Line music, and Vibe introduced Karaoke mode and Samsung UHDTV introduced dialogue enhancement. We targeted these early multimedia related markets and defined our vocal and accompaniment separation model criteria in terms of robustness, quality, and cost.

First, the model should have robustness over different types of audio signals. Typical pop music pieces nowadays have an integrated loudness [1] of around -10 LUFS [1], TV programs have -25~-23 LUFS following the international standards such as [2], and some YouTube programs have below -40 LUFS. If a separation model is trained for pop music, it may not provide the best performance for the TV programs. The model can be more robust against loudness differences by a set of proper augmentations, but not the best because it is a regression model.

Second, the model should provide equivalent or superior sound quality compared to the state-of-the-art technologies. The objective measure of SDR (Source to Distortion Ratio) and SIR (Source to Interference Ratio) [3] shows all the technologies promising but some of the resulting audio signals have artifacts for music separation of coloration, fluctuation, and additive noises.

Third, the model should be computationally efficient enough to be implemented on a wide range of consumer electronics products from UHDTV to smartphones. For the streaming service providers with a server-client architecture, the computational efficiency is also important from the viewpoint of the operation cost. It is noteworthy that the Spotify adds over 40,000 new songs every day. [4]

In this paper, observations on the three major state-of-the-art source separation models, the Open-Unmix [5], Demucs [6], and Spleeter [7], based on these three criteria are explained in section 2, the proposed model from the observations is introduced in chapter 3, the experiments to evaluate the proposed system are explained in chapter 4, and finally conclusion is made in chapter 5.

## 2. OBSERVATIONS AND DESIGN PRINCIPLES

### 2.1. Robustness in Loudness

**Table 1** SDR of the voice separated signal at -15, -30, -45 LUFS

| Model | Vocal Separation SDR | | |
|---|---|---|---|
| | -15 LUFS | -30 LUFS | -45 LUFS |
| Open-Unmix | 6.24 | 5.36 | 0.21 |
| Demucs | 6.86 | 6.86 | 6.86 |
| Spleeter | 6.69 | 5.50 | 1.25 |

To verify whether the state-of-the-art separation models are robust against the loudness difference of the input program, we measured vocal SDR of each model for the MUSDB18 [8] after the loudness normalization with target loudnesses at -15, -30, and -45 [LUFS]. In the loudness normalization, the integrated loudness of each excerpt was measured following ITU-R Recommendation BS.1770-3. [1]

As shown in Table 1, the vocal separation SDR of the Open-Unmix and Spleeter model decreases as the loudness of the input mixture decreases. The Demucs, on the other hand, has stable performance due to standardization logic pair in pre-processing and post-processing.

## 2.2. Sound Quality Enhancement for Real Application

While both SDR and SIR are popular measures in the audio source separation, higher SDR and SIR do not guarantee better perceptual sound quality. After a set of benchmark studies, we found the following structure-to-quality relationships.

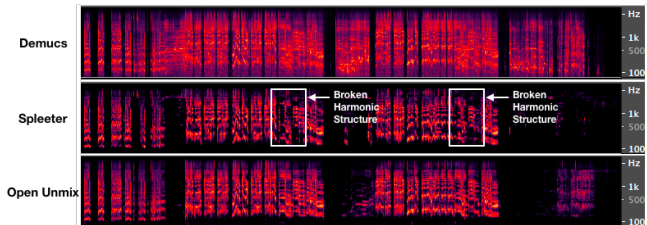

**Figure 1** A vocal separated signal example. (David Bowie – Space Oddity)

*2.2.1. Time domain model vs. frequency domain model*
Besides the interference of the unwanted signal, the time domain model such as Demucs and Wave-U-Net [9] brings more wideband noise than the frequency domain model such as Open-Unmix or Spleeter does as shown in Figure 1. A possible explanation may be that the regression error by the time domain model brings sample-wise discontinuities while that by the frequency domain model brings STFT frame-wise discontinuities. Another possible explanation may be that the time domain model does not have an "overlap-and-add" method, which smoothly interpolates the abrupt discontinuity caused by time-varying signal processing in the frequency model. For the target market of Karaoke, we found frequency domain model more stable from the viewpoint of the noise.

*2.2.2. Kernel design*
When we compare two frequency domain models of Open-Unmix and Spleeter, the Spleeter output signal often has a broken harmonic structure also shown in Figure 1. It can be explained by that the two-dimensional convolution network used in the Spleeter for a frequency component misses the useful information at lower or higher frequency components which is out of the kernel range. It may be resolved by using bigger size of the kernel, but it increases model complexity. For the stability of the output signal, we found a one-dimensional model more appropriate and the Open-Unmix was chosen as a baseline system.

*2.2.3. Interference*
Internal benchmark showed that the Open-Unmix, the baseline, has lower SIR comparing to the Demucs and Spleeter. The signal analysis block of the Open-Unmix is three LSTMs with a skip connection and it may be not enough to remove the interference of the unwanted signal. For better voice activity detection and feature extraction, we used a more sophisticated combination of gating components, such as gated convolutions [10], highway networks [11], GRU (gated recurrent networks) [12].

Another way to reduce the interference is by warping the mask in a more conservative way. As the soft mask in the Open-Unmix is trained by regression, the input signal with a lower mask in a time-frequency bin is more likely to have a strong interference. Here, mask=1 means extracting the signal, which is opposite to literal negative interpretation. By warping the mask with a nonlinear function, the interference can be reduced. Figure 2 shows examples of the non-linear mask warping functions – power warping, tangent sigmoid warping, and exponent warping.

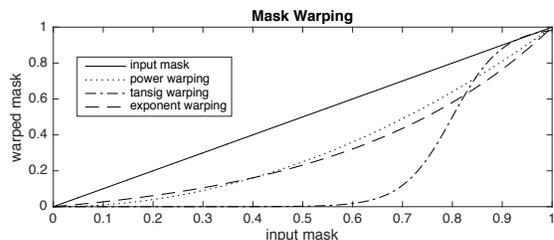

**Figure 2** Examples of mask warping functions

## 2.3. Computational Efficiency

Computational efficiency is as critical as sound quality in a real-world application. As a benchmark test, we measured the average inference time per second input on the Open-Unmix, Demucs, and Spleeter running on a GPU server and a mobile device. When we measured on a GPU server, they consumed 1.8~8.7 [msec] processing time for one second input. When we implemented on a mobile device, Open-Unmix and Spleeter consumed 94.6 and 23.32 [msec], respectively. Demucs was not able to be implemented on the mobile device because the model size is beyond capacity of the testing mobile device. The measured processing times are shown in Table 5 in 4.2.

## 3. PROPOSED SEPARATION SYSTEM

### 3.1. System Architecture

Figure 3 shows the system architecture of the proposed separation system. A loudness normalization and de-normalization pair is used for the robustness against the loudness differences of mixtures, a Gated CBHG is designed for better feature analysis and voiced/unvoiced detection, a mask warping is added to reduce the interference by the unwanted sources.

*3.1.1. Loudness normalization and de-normalization pair*
The input mixture $m(n)$ is normalized to a target loudness and the loudness normalized input signal $m_{LN}(n)$ is used as an input of the deep learning model. After the source separation, the separated model output signal $\hat{s}_{LN}(n)$ is de-normalized using the loudness normalization gain $g_{LN}$ to get the separated system output signal $\hat{s}(n)$. Here, the

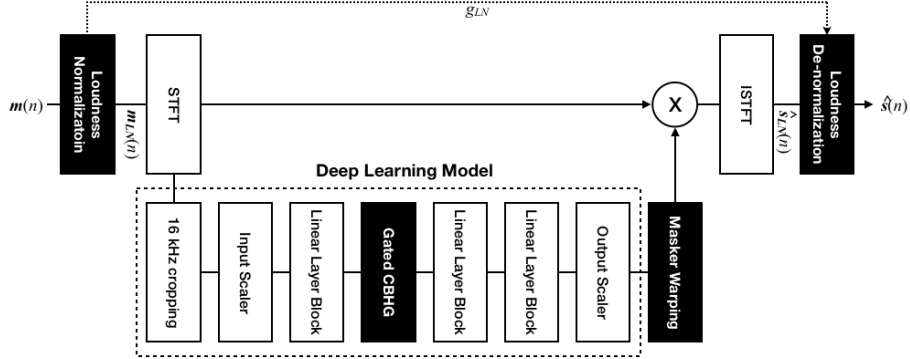
**Figure 3** Architecture of the Proposed System

normalization gain is defined as $g_{LN} = 10^{(L_T - L_I)/20}$ for target loudness $L_T$ and integrated loudness $L_I$. The integrated loudness is calculated following [1] but other loudness models can also be used.

*3.1.2. Deep Learning Model with Gated CBHG*

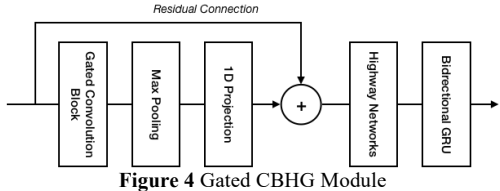
**Figure 4** Gated CBHG Module

The deep learning model in Figure 3 is identical to the Open-Unmix except that a CBHG is used instead of the 3-layer LSTMs. The CBHG module was first introduced in [13] for character level neural machine translation. In Tacotron[14], a speech synthesis model, the CBHG module was also used for text encoding and transformation from melspectrogram to linear spectrogram. We deployed the CBHG module to analyze the context of the given mixture spectrogram. Then, we added gated linear units [10] in the convolution layers of the CBHG to improve voice/unvoiced detecting ability.

The dimension of the input and output of the CBHG module is 512. The convolution block has eight convolution layers and each kernel width is 1 to 8 with a channel size of 256. The max pooling is performed along the time axis with a pooling size of 2. After the one-dimensional projection, the output dimension becomes 512 for the residual connection. The dimension of layers inside highway networks is 512. In the bidirectional GRU, the hidden size in each direction is 256 and the final output size is 512.

*3.1.3. Mask Warping*

For the mask warping, we chose a simple power function of $f(x) = x^\alpha$ where $\alpha$ is a warping intensity ratio.

### 3.2. Training Details

We trained two models – one model for voice extraction and the other for accompaniment extraction. The models were trained with MUSDB18 and extra datasets of 3000 private music and public speech datasets (LibriSpeech [15] and KsponSpeech [16]). Each audio segment for training was created considering the loudness normalization and augmentation by following steps.
1) Randomly choose one voice source and adjust the loudness to 0 LUFS.
2) Randomly choose three non-voice source and adjust the random loudness value between -12LUFS to 12LUFS.
3) Mix loudness adjusted sources.

In the training, a batch has 80 audio segments, a mean square error was used as a loss function, and an Adam optimizer [17] with a learning rate 1e-3 and a weight decay 1e-5 was used. The learning rate was reduced by the ReduceLROnPlateu scheduler in the PyTorch framework with a decay gamma of 0.9, a decay patience of 140, and a cooldown of 10. During the training, the loudness normalization pair were used with the target loudness $L_T$ of -13 LUFS and the mask warping block was bypassed.

### 3.3. Inference

The loudness normalization pair should be used with the target loudness $L_T$ of -13 LUFS as the model is optimized for the -13LUFS programs and the mask warping block should be used.

## 4. EXPERIMENT

### 4.1. Subjective Evaluation

To evaluate the sound quality of the proposed system, two sets of listening tests were carried out by ten listeners following ITU-R BS.1534-3 [18] except for the use of a hidden reference and anchor. We defined the input mix signal of the system as a "mix reference" and the listeners were asked to evaluate basic audio quality with how much each signal under test is close to the listener's imaginary voice or accompaniment in the mix reference.

The test sets are (1) accompaniment separation from music for Karaoke and (2) voice separation from movie/TV

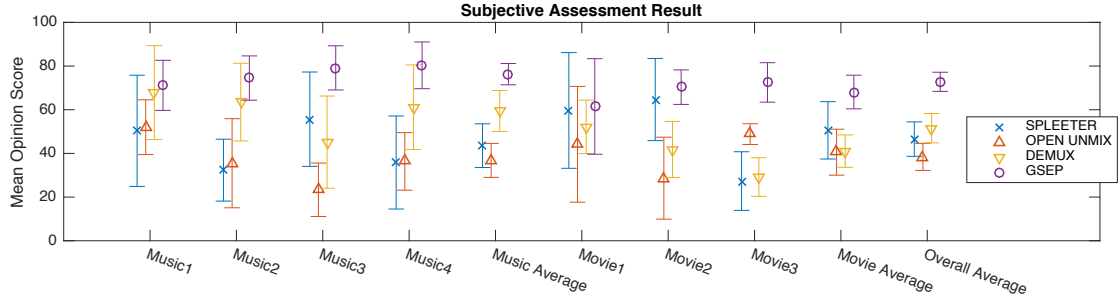

**Figure 5** Subjective Assessment Result (95% confidence interval)

for dialogue enhancement. The excerpts used in test 1 are described in Table 2 and those used in test 2 are described in Table 3. As shown in Table 2 and 3, the test excerpts are selected from the real-world pieces and programs instead of MUSDB18. As the system under tests, four conditions were compared – GSEP, Demucs, Open-Unmix with Wiener filter, Spleeter with Wiener Filter. The subjective assessment result is shown in Figure 5 and it was verified that the GSEP provides better sound quality showing the highest MOS value for all seven excerpts. Especially for the movie excerpt 2 and 3, the GSEP showed significantly better quality. The main reasons are that the GSEP removed the crowd noise for movie 3 and kept secondary dialogues, which were relatively lower in loudness than the primary dialogue, for movie 2 while the other models did not.

**Table 2** Music test excerpts for test 1 for Karaoke

| ID | Title | Scene description |
|---|---|---|
| 1 | Cardi B | Heavy bass hiphop with male & female vocal |
| 2 | Red Velvet | K-Pop with female solo & chorus |
| 3 | Damien Rice | Acoustic with male vocal, guitar, and strings |
| 4 | Imagine Dragon | Pop with male vocal & chorus |

**Table 3** Movie/TV test excerpts for test 2 for dialogue enhancement

| ID | Title | Scene descriptions |
|---|---|---|
| 1 | House of Cards | Dialogue, court crowd noise, BGM |
| 2 | Marriage Story | Primary and secondary dialogues, BGM |
| 3 | Sport Event | Caster dialogue, stadium crowd noise, BGM |

### 4.2. Objective Evaluation

For the objective evaluation for the separation performance, the SDR and SIR of GSEP were compared with those of the other models as shown in Table 4. Some of the values of the compared models were publicly reported and the others were measured by the authors using museval software [19]. In Table 4, GSEP and GSEP+WF represent the proposed model without Wiener filter and with Wiener filter, respectively. As shown in Table 4, it was verified that the proposed system satisfies the equivalent or higher separation quality in objective measures showing that GSEP has a competitive SDR and SIR even without Wiener filter and GSEP+WF has the highest vocal SDR, vocal SIR, and accompaniment SIR and third highest accompaniment SDR.

For the computational efficiency, the processing time of GSEP was measured and compared. As shown in Table 5, it was verified that the proposed system has the lowest processing time on GPU server and second lowest on mobile device.

**Table 4** Objective evaluation result

| Model | Vocal | | Accompaniment | |
|---|---|---|---|---|
| | SDR[1] | SIR[1] | SDR[1] | SIR[1] |
| MMDenseLSTM+WF[4] | 7.16[1] | 16.49[1] | **13.73**[1] | 18.50[1] |
| Demucs | 7.05[2] | 13.94[2] | 13.37[3] | 17.95[3] |
| Open-Unmix+WF[4] | 6.32[2] | 13.33[2] | 12.73[3] | 18.47[3] |
| Spleeter+WF[4] | 6.86[2] | 15.86[2] | 12.70[3] | 18.88[3] |
| **GSEP**[5] | 6.98 | 15.38 | 13.28 | 17.65 |
| **GSEP+WF**[5] | **7.24** | **17.62** | 13.30 | **18.98** |

1) Shown during the SiSEC 2018. [19]
2) Reported in the papers. [5,6,7]
3) Publicly unavailable and measured using [19] by the Authors.
4) Wiener filter was used for the MMDenseLSTM, Open-Unmix, and Spleeter models.
5) The intensity ratio $\alpha$ is 1.4.

**Table 5** Average inference time per second

| Model | processing time[1] per second [ms] | | Model Size |
|---|---|---|---|
| | Server[2] | Mobile[3] | |
| Open-Unmix[6] | 3.82 | 94.60 | **35 MB** |
| Demucs | 8.69 | N/A[4] | 2.59 GB |
| Spleeter[6] | 1.81[5] | **23.32**[5] | 37 MB |
| **GSEP**[6] | **1.49** | 39.09 | 96 MB |

1) Processing time is measured fifty times for a three-minute input signal and averaged among the forty fastest measured time.
2) Server specification: Intel Xeon Gold 5120, NVDIA V100, Ubuntu 16.04, PyTorch 1.6.
3) Mobile device specification: Samsung Galaxy 9, Android 9, PyTorch 1.6.
4) The model size of the Demucs is beyond the capacity of the testing mobile device.
5) Spleeter models on the server and mobile was re-implemented in PyTorch 1.6 by the authors
6) Open-Unmix, Spleeter, and GSEP was implemented without Wiener filter.

### 5. CONCLUSION AND FUTURE WORKS

GSEP was designed with a set of principles in terms of robustness, quality, and cost for the Karaoke and dialogue enhancement system which are (1) robustness against the loudness differences, (2) equivalent or better sound quality, and (3) low computational complexity to support a wide range of the market needs. For the principles, we implemented the GSEP system with loudness normalization, gated CBHG, and mask warping. It was verified that all the principles were satisfied through both objective and subjective assessments.

As future works, we will add fourth design principle of "live streaming support" and study on the unidirectional RNN or very short input audio segmentation.